\documentclass[runningheads]{llncs}

\usepackage{comment}
\usepackage{float}

\usepackage{amssymb}
\usepackage{amsmath}
\usepackage{graphicx}
\usepackage{xcolor}
\usepackage{url}
\usepackage{hyperref}
\usepackage{tabularx}
\usepackage{subcaption}
\usepackage{makecell}
\usepackage{cite}

\usepackage[colorinlistoftodos]{todonotes}
\urldef{\mailsb}\path|jan.valdman@utia.cas.cz |

\providecommand{\keywords}[1]{\textbf{\textit{Keywords:}} #1}

\newcommand{\R}{\mathbb R}
\newcommand{\N}{\mathbb N}

\newcommand{\m}{m}
\newcommand{\p}{p}

\newcommand{\dxdy}{\mathrm{d}\boldsymbol{\mathrm{x}}}

\newcommand{\hpindx}{B(\mathcal{T},\p)}
\newcommand{\indx}{C(\mathcal{T},\p)}
\newcommand{\signs}{S(\mathcal{T},\p)}
\newcommand{\Mk}{M(T_k)}
\newcommand{\Kk}{K(T_k)}
\newcommand{\Kref}{K^{ref}}
\newcommand{\Mref}{M^{ref}}
\newcommand{\Nm}{N_{\m}}
\newcommand{\Nmx}{\Nm(\xi)}
\newcommand{\nb}{n_{\p}}
\newcommand{\nbref}{n_{\p,ref}}
\newcommand{\Nig}{N_i^{(g)}}
\newcommand{\Njg}{N_j^{(g)}}
\newcommand{\Nmg}{N_m^{(g)}}
\newcommand{\SpT}{S^p(\mathcal{T})}

\newcommand{\nt}{|\mathcal{T}|}
\newcommand{\nn}{|\mathcal{N}|}
\newcommand{\ned}{|\mathcal{E}|}

\newcommand\Tr{T_{ref}}

\newcolumntype{C}{>{\centering\arraybackslash}X}
\newcolumntype{L}[1]{>{\RaggedRight\hspace{0pt}}p{#1}}
\newcolumntype{R}[1]{>{\RaggedLeft\hspace{0pt}}p{#1}}
\newcolumntype{d}{>{\hsize=1\hsize}X}
\newcolumntype{s}{>{\hsize=.5\hsize}Y}
\newcolumntype{z}{>{\hsize=1.4\hsize}Y}

\title{MATLAB implementation of hp finite elements on rectangles
\thanks{A. Moskovka was supported by the R\&D project 8J21AT001 Model Reduction and Optimal Control in Thermomechanics. J. Valdman announces the support
of the Czech Science Foundation (GACR) through the GF21-06569K grant Scales and shapes in continuum thermomechanics. 
} 
}
\titlerunning{MATLAB Implementation of C1 finite elements}

\author{Alexej Moskovka\inst{1}\orcidID{0000-0003-0091-151X} \and Jan Valdman\inst{2,3}\orcidID{0000-0002-6081-5362} }
\authorrunning{Alexej Moskovka, Jan Valdman}
\institute{Department of Mathematics, Faculty of Applied Sciences,
University of West Bohemia, Technick\' a 8, 30100 Plze\v n, Czechia \\
\and
Institute of Mathematics, Faculty of Science, University of South Bohemia, 
Brani\v sovsk\' a 31, 37005~\v{C}esk\'{e}~Bud\v{e}jovice, Czech Republic \and
The Czech Academy of Sciences, 
Institute of Information Theory and Automation, 
Pod vod\'{a}renskou v\v{e}\v{z}\'{\i}~4, 18208~Praha~8, Czech Republic \\
\email{jan.valdman@utia.cas.cz}
}

\begin{document}
\maketitle
\begin{abstract}
A simple MATLAB implementation of hierarchical shape functions on 2D rectangles is explained and available for download. Global shape functions are ordered for a given polynomial degree according to the indices of the nodes, edges, or elements to which they belong. For a uniform p-refinement, the hierarchical structure enables an effective assembly of mass and stiffness matrices. A solution of a boundary value problem is approximated 
for various levels of uniform h and p refinements. 

\end{abstract}

\section{Introduction}
hp-FEM is a numerical method for solving partial differential equations based on piecewise polynomial approximations that employ elements of variable size ($h$) and degree of polynomial ($\p$). The origins of hp-FEM date back to the works of Ivo Babuška and his coauthors in the early 1980s (e.g. \cite{BSK, BabuskaGuo}) who discovered that the finite element method converges exponentially fast when the mesh is refined using a suitable combination of h-refinements (dividing elements into smaller ones) and p-refinements (increasing their polynomial degree). 
Many books (e.g. \cite{BabuskaSzabo1, BabuskaSzabo2, Demkowicz, Solin_HOFEM}) have been written explaining the methodology of hp-FEM accompanied by software codes \cite{Schoeberl, SolinHermes} in C++.

Implementing hierarchical shape functions, particularly in the case of hp adaptivity, is not straightforward, and special data structures are needed \cite{Bangerth, DemkowiczOden}. A recent MATLAB contribution \cite{MOOAFEM} provides an object-oriented approach to implement hp-FEM with adaptive h refinement. Our focus is on a simple MATLAB implementation directly based on \cite{BabuskaSzabo2}. We provide eight examples that demonstrate the basics of hp-FEM assemblies, including:
\begin{itemize}
\item[$\bullet$] constructions of basis functions and their isoparametric transformations to general quadrilaterals (Sec. \ref{sec:2}),
\item[$\bullet$] the ordering of global shape functions using indexing matrices (Sec. \ref{sec:3}),
\item[$\bullet$] assemblies of the mass and stiffness matrices (Sec. \ref{sec:4}),
\item[$\bullet$] solution of a particular diffusion-reaction boundary value problem using uniform h and p refinements (Sec. \ref{sec:5}).
\end{itemize}

\section{Hierarchic shape functions} \label{sec:2}
We consider the basis functions for the dimensions of space $d \in \{1,2\}$ (see \cite{BabuskaSzabo2}). For a reference element $ \Tr = [-1,1]^d$ and $p \in \N$ we denote by 
\begin{equation}
S^p(\Tr)    
\end{equation}
the space of polynomials of degree $\p$ defined on $\Tr$.
The basis functions that span the space are called shape functions. We define them using Legendre polynomials for $x \in [-1,1]$:
\begin{gather} \label{legendre}
    \begin{split}
        & P_0(x) = 1 \, , \quad  P_1(x) = x \, , \\
        & P_{n+1}(x) = \frac{(2n+1)\,x\,P_n(x) - n\,P_{n-1}(x)}{n+1} \, , \qquad n \geq 1 \, .
    \end{split}
\end{gather}
\subsubsection{Hierarchic shape functions on $T_{ref} = [-1,1]$} 
\begin{figure}[H]
    \centering
    \includegraphics[width=0.95\textwidth]{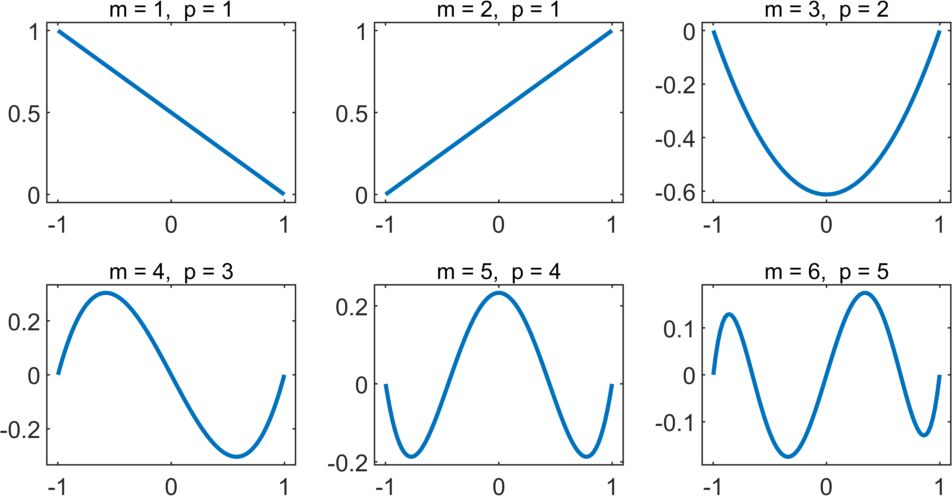}
    \caption{The hierarchic shape basis functions $\Nmx, \; m=1,\dots,6$, where $\p$ is the corresponding polynomial degree.}
    \label{fig:i_1_9}
    \vspace{-6mm}
\end{figure}

are functions 
$\Nmx : T_{ref} \rightarrow \R, \; \m \in \N$ defined using \eqref{legendre} as:
\begin{equation} \label{hp_basis_1D}
    \begin{split}
        N_1(\xi) &= \frac{1-\xi}{2} \, ,  \quad N_2(\xi) = \frac{1+\xi}{2} \, , \\
        \Nmx &= \frac{1}{\sqrt{2(2\m-3)}} \big(P_{\m-1}(\xi) - P_{\m-3}(\xi)\big) \, , \qquad \m \geq 3 \, .
    \end{split}
\end{equation}
All $\Nmx, \; m \geq 3$ vanishes at the endpoints of $\Tr$.
\begin{example}
The first hierarchic shape functions are shown in Fig \ref{fig:i_1_9}
and the pictures can be reproduced by the script
\begin{verbatim}
  example1_draw_hp_basis_1D
\end{verbatim}
\end{example}

\subsection{Hierarchic shape functions on $\Tr = [-1,1]^2$}
For $\p \in \N$ we define the trunk space $S^p(\Tr)$ spanned by polynomials $\xi^i \eta^j,$ where $i,j \in \N_0$ satisfies $i+j \leq p$, supplemented by the polynomial $\xi \eta$ for $\p = 1$ and the polynomials $\xi^p \eta, \, \xi \eta^p$ for $p \geq 2$. Its dimension is given by
\begin{equation} \label{dimSp}
    \nbref = \dim(S^p(\Tr)) =
    \begin{cases}
        4p \, , \qquad &p \leq 3 \\
        4p + (p-2)(p-3)/2 \, , \quad &p \geq 4 \, .
    \end{cases}
\end{equation}
There are three types of 2D shape function: nodal (Q1), edge, and bubble (sometimes called internal). 
The nodal shape functions that span the space $S^1(\Tr)$ are defined as follows:
\begin{equation} \label{hp_basis_2D_nodal}
    \begin{split}
        N_1(\xi,\eta) = \frac{1}{4} (1-\xi)(1-\eta) \, , \qquad  N_2(\xi,\eta) = \frac{1}{4} (1+\xi)(1-\eta) \, , \\
        N_3(\xi,\eta) = \frac{1}{4} (1+\xi)(1+\eta) \, , \qquad         N_4(\xi,\eta) = \frac{1}{4} (1-\xi)(1+\eta) \, .
    \end{split}
\end{equation}
The function of the $i$-th nodal shape is equal to one in the $i$-th node of $\Tr$ and vanishes in other nodes. 
Edge shape functions are constructed by multiplying one-dimensional shape functions $\Nmx, \, \m \geq 3$ from \eqref{hp_basis_1D} by linear blending functions. We define
$  \phi_{\p}(x) = N_{\p+1}(x), \, \p \geq 2,$
and the edge shape functions 
by
\begin{equation} \label{hp_basis_2D_edge}
    \begin{split}
        N_{\p}^{(1)}(\xi,\eta) &= \frac{1}{2} (1 - \eta) \, \phi_{\p}(\xi) \, , \qquad \;\;\;
        N_{\p}^{(2)}(\xi,\eta) = \frac{1}{2} (1 + \xi) \, \phi_{\p}(\eta) \, , \\
        N_{\p}^{(3)}(\xi,\eta) &= \frac{1}{2} (1 + \eta) \, \phi_{\p}(-\xi) \, , \qquad 
        N_{\p}^{(4)}(\xi,\eta) = \frac{1}{2} (1 - \xi) \, \phi_{\p}(-\eta) \, .
    \end{split}
\end{equation}
For $j \in \{1,2,3,4\}$, the restriction of $N_{\p}^{(j)}$ on the $j$-th edge is equal to the corresponding one-dimensional edge shape function of the $\p$-th degree, and it vanishes along the other edges.
Bubble-shaped functions are defined as
\begin{equation} \label{hp_basis_2D_bubble}
    N_{\p}^{\beta}(\xi,\eta) = \phi_{p-(\beta+1)}(\xi) \, \phi_{\beta+1}(\eta) \, , \qquad 1 \leq \beta \leq p-3 \, , \quad \p \geq 4
\end{equation}
and any of them attains zero values on all edges. Table \ref{tab:hp_numbers} shows the number of shape functions in $\Tr$ for $1 \leq p \leq 7$.
\begin{table}[H]
    \centering
    \resizebox{0.85\columnwidth}{!}{
    \begin{tabularx}{\textwidth}
    {C |C |C |C | C}
   polynomial degree $\p$ & $\#$ of nodal functions & $\#$ of edge functions & $\#$ of bubble functions & $\#$ of all functions  
   \\
     \hline 
 1 & 4 & 0 & 0 & 4  \\
 2 & 4 & 4 & 0 & 8  \\
 3 & 4 & 8 & 0 & 12  \\
 4 & 4 & 12 & 1 & 17  \\
 5 & 4 & 16 & 3 & 23  \\
 6 & 4 & 20 & 6 & 30  \\
 7 & 4 & 24 & 10 & 38  \\
    \end{tabularx}}
    \vspace{2mm}
    \caption{The numbers of shape functions.} \label{tab:hp_numbers}
    \vspace{-0.8cm}
\end{table}
\subsubsection{Local indexing}
The shape functions of the $\p$ th degree in $\Tr$
are ordered by a unique index $\m \in \N$ given by
\begin{equation}
    m =
    \begin{cases}
        4(p-1) + s \, , \qquad &\mbox{for} \;\; p \leq 4 \, , \\
        4(p-1) + (p-3)(p-4)/2 + s \, , \qquad &\mbox{for} \;\; p \geq 5 \, ,
    \end{cases}
\end{equation}
\begin{description}
\item where for $\p = 1$: \;\,$s$ is the index of a node $i \in \{1, 2, 3, 4 \},$ 
\item  \hspace*{9mm} for $\p \geq 2$: \;\,$s$ is the index of an edge $j \in \{1, 2, 3, 4 \}$,
\item \hspace*{9mm} for $p \geq 4$: \;\:$s = 4+\beta,$ where $\beta$ is the local index of a bubble
function \eqref{hp_basis_2D_bubble}.
\end{description}

\begin{figure}
    \centering
    \includegraphics[width=\textwidth]{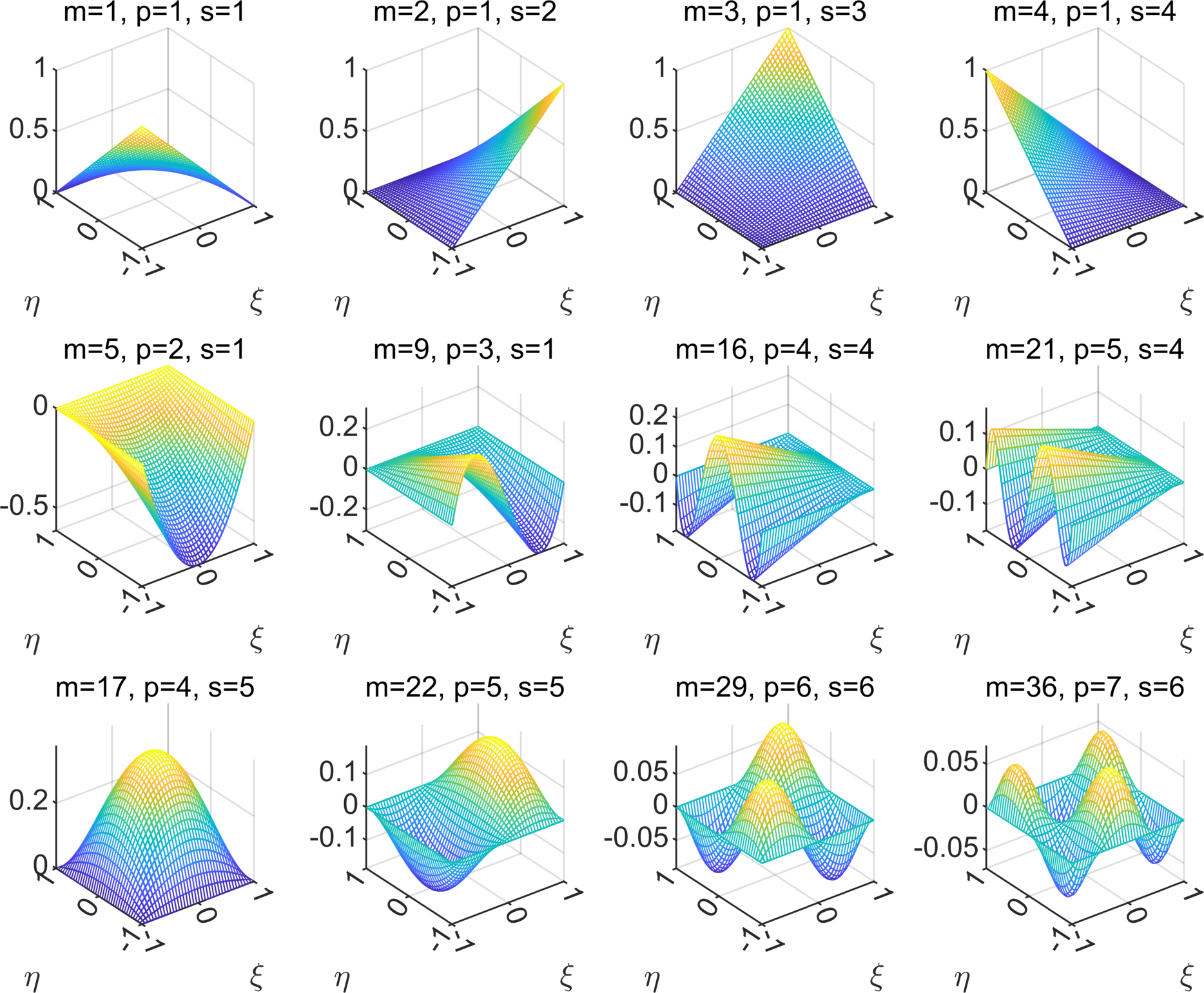}
    \caption{Examples of nodal (the top row), edge (the middle row) and bubble (the bottom row) shape functions.}
    \label{fig:hp_basis_2D}
\end{figure}

\begin{example}
Several shape functions are depicted in Fig. \ref{fig:hp_basis_2D} and can be reproduced by the script
\begin{verbatim}
  example2_draw_hp_basis_2D
\end{verbatim}
The polynomial degree $\p$ and the local index $s$ are evaluated by the function
\verb+[s,p] = shapeindx(m)+.
\end{example}
\begin{figure}
\centering
\includegraphics[width=0.8\textwidth]{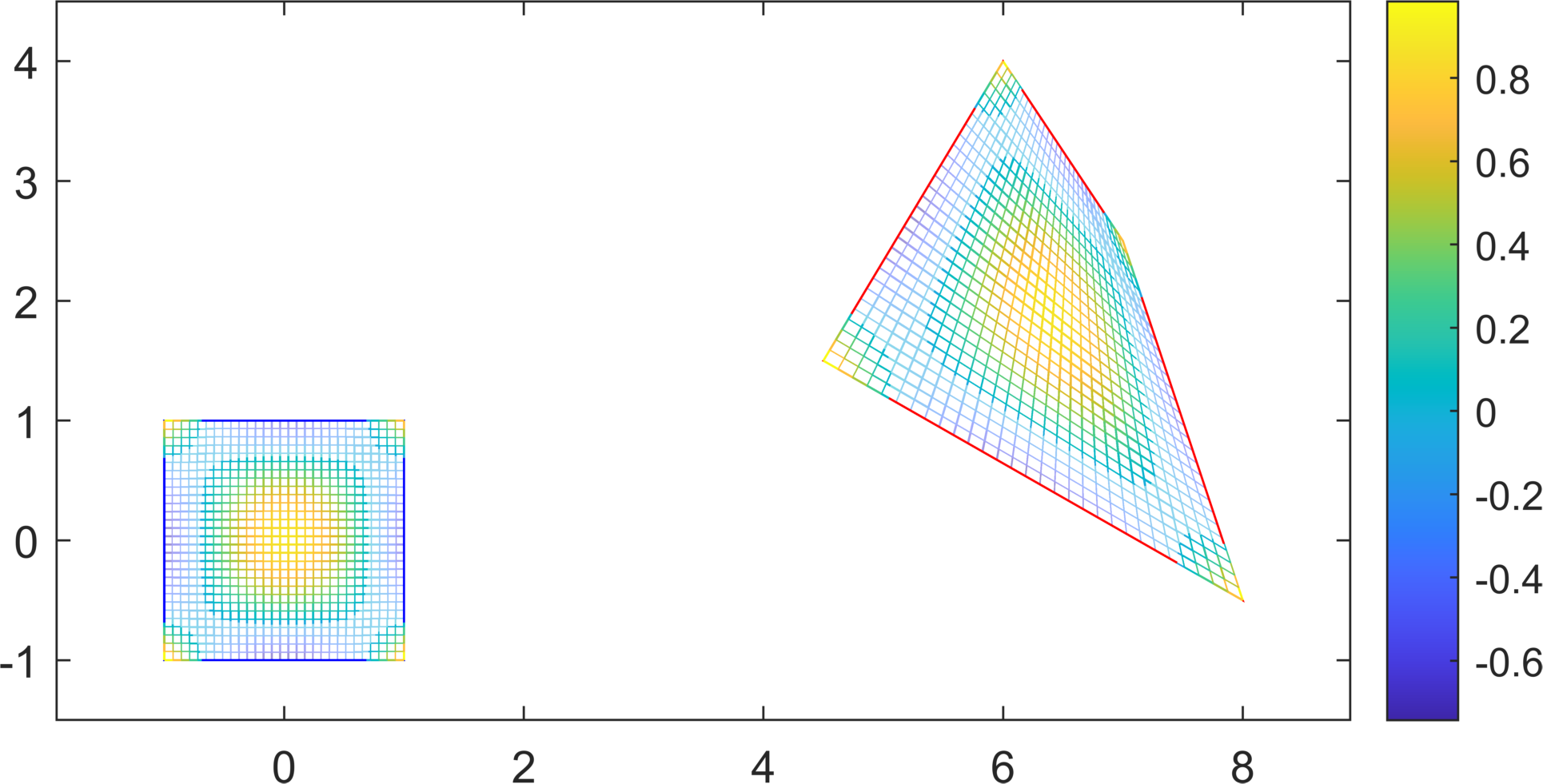}
\caption{The isoparametric transformation of $\Tr$ to a quadrilateral $T$ and the approximation of the function $\cos{(\frac{3\pi}{4}\xi)}\cos{(\frac{3\pi}{4}\eta)}$ for $p=4$ .}
\label{fig:isoparametric}
\end{figure}

\vspace{-3mm}
\subsubsection{Mapping from $\Tr$ to a quadrilateral $T$}
Transformation of a reference element $\Tr$ to a quadrilateral $T$ is performed by the isoparametric mapping $Q : \Tr \rightarrow T$ defined as $
        (x,y)(\xi,\eta) = Q(\xi,\eta),$
where        
\begin{equation} \label{isoparametric_mapping}
    \begin{split}
        Q(\xi,\eta) = \Big( \sum_{i=1}^4 X_i \, N_i(\xi,\eta) \, , \sum_{i=1}^4 Y_i \, N_i(\xi,\eta) \Big) \, ,
    \end{split}
\end{equation}
and $(X_i,Y_i), \; i \in \{1,2,3,4\}$ are the coordinates of the $i$-th node of $T$. For a given $p \in \N$ we denote by
\begin{equation}
    S^p(T)
\end{equation}
the space of functions spanned by $N_m\big(Q^{-1}(x,y)\big)$,  where $N_m \in S^p(\Tr)$.

\begin{example} \label{ex:3}
The transformation of $\Tr$ into a quadrilateral $T$ given by the nodes with coordinates 
$$(9/2,3/2), \, (8,-1/2), \, (7,5/2), \, (6,4)$$ is shown in Fig. \ref{fig:isoparametric} and can be reproduced by the script
\begin{verbatim}
  example3_isoparametric_transformation
\end{verbatim}
Additionally, it visualizes the approximation for $\p = 4$ of the function 
$f(\xi,\eta) = \cos{(\frac{3\pi}{4}\xi)}\cos{(\frac{3\pi}{4}\eta)}, \, (\xi,\eta) \in \Tr$ and its transformation to quadrilateral $T$. Visualization is implemented by the function \verb+visualize_hp(mesh,u)+, where \verb+mesh+ specifices the mesh properties  and \verb+u+ is a vector of the coefficients of function $f$ in $S^4(\Tr)$ basis. According to Table \ref{tab:hp_numbers}, there are 17 coefficients. 
\end{example}

\section{Global shape functions} \label{sec:3}
A domain $\Omega \subset \R^2$ is approximated by a 
rectangulation $\mathcal{T}$ into closed elements (quadrilaterals). We denote by $\mathcal{N}, \mathcal{E}$ and $\mathcal{T}$ the sets of nodes, edges, and elements, respectively, and by $\nn, \ned$ and $\nt$ their sizes. For a given $p \in \N$ we define it by
\begin{equation}
    \SpT
\end{equation}
the space of all global shape functions on $\mathcal{T}$ and by $\nb$ its dimension given by
\begin{equation}
    \nb =
    \begin{cases}
        \nn + (p-1) \, \ned \, , \qquad &p \leq 3 \, , \\
        \nn + (p-1) \, \ned + \frac{1}{2}(p-2)(p-3) \, \nt \, , &p \geq 4 \, .
    \end{cases}
\end{equation}
We denote by $\Nmg, \, 1 \leq m \leq \nb$ the $m$-th global shape function defined by its restrictions on elements $T_k \in \mathcal{T}, \, 1 \leq k \leq \nt$ in the following way:
\begin{description}
\item $\Nmg$ is a nodal shape function corresponding to the $i$-th node:
If $T_k$ is adjacent to the $i$-th node, then $\Nmg\big|_{T_k} = \tilde{N}_{l,k}$, where $\tilde{N}_{l,k}$ is the $l$-th local nodal shape function on $T_k$ which is equal to one in the $i$-th node. Otherwise, $\Nmg = 0$.
\vspace{1mm}
\item $\Nmg$ is an edge shape function corresponding to the $j$-th edge:
If $T_k$ is adjacent to the $j$-th edge, then $\Nmg\big|_{T_k} = \tilde{N}_{l,k}$, where $\tilde{N}_{l,k}$ is the $l$-th local edge shape function on $T_k$ whose restriction on the $j$-th edge is the corresponding edge shape function in 1D. Otherwise, $\Nmg = 0$.
\vspace{1mm}
\item $\Nmg$ is a bubble shape function corresponding to the $k$-th element: \\ $\Nmg\big|_{T_k} = \tilde{N}_{l,k}$, where $\tilde{N}_{l,k}$ is the corresponding $l$-th local bubble shape function on the $k$-th element. Otherwise $\Nmg = 0$.
\end{description}

\vspace{-10mm}
\begin{figure}[H]
\centering
\hspace{0.1\textwidth}
\begin{minipage}[c]{0.95\textwidth}
\includegraphics[width=\textwidth]{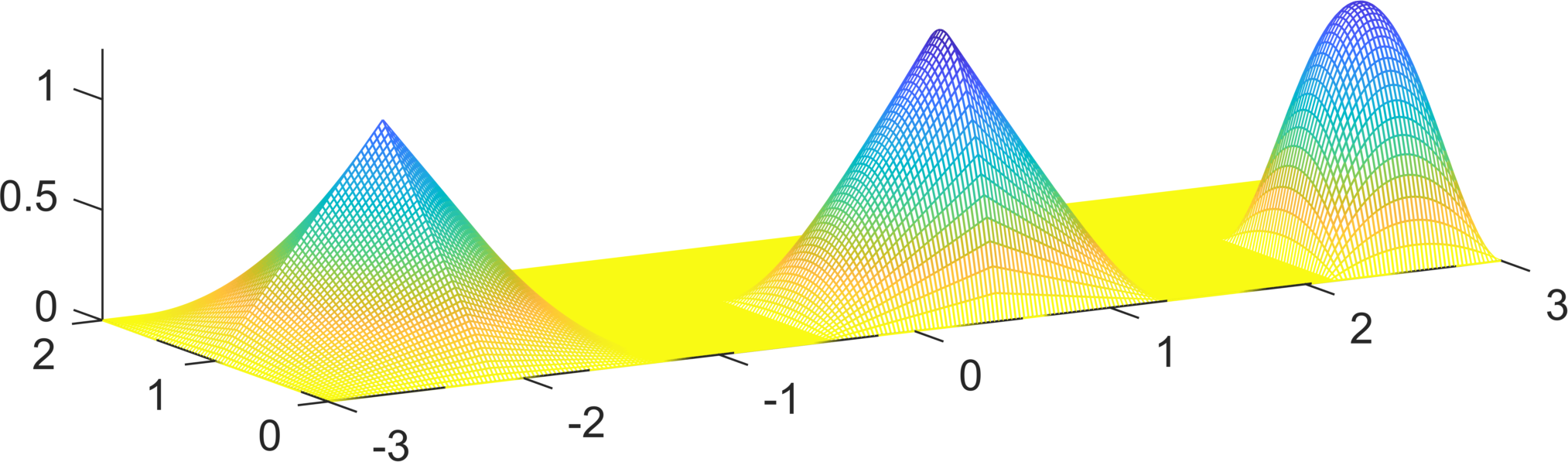}
\vspace{-4mm}
\end{minipage}
\begin{minipage}[c]{0.85\textwidth}
\hspace*{3mm}
\includegraphics[width=\textwidth]{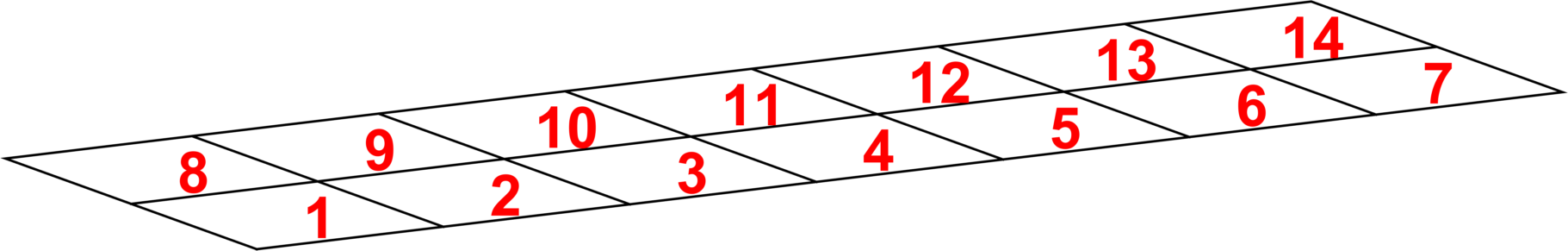}
\end{minipage}
\:\:\:
\caption{Function $u \in \SpT \in $ of \eqref{u_global} and the underlying rectangular mesh with indices of elements.}
\label{fig:nodal_edge_bubble}
\vspace{-4mm}
\end{figure}

\begin{example}
We assume a rectangulation $\mathcal{T}$ of $\Omega = (-3,3) \times (0,2)$  with $\nn = 24$, $\ned = 37$, $\nt = 14$, $\nb = 149$ and the function $u \in S^4(\mathcal{T})$ defined as
\begin{equation} \label{u_global}
u(x,y) = N_{10}^{(g)}(x,y) -2 N_{34}^{(g)}(x,y) -2 N_{142}^{(g)}(x,y) \, , \qquad (x,y) \in \Omega
\end{equation}
shown in Fig. \ref{fig:nodal_edge_bubble}. The nodal function $N_{10}^{(g)}$ corresponds to the node adjacent to $T_1, T_2, T_8, T_9$,  the edge function $N_{34}^{(g)}$ to the edge adjacent to $T_4, T_5$, and the bubble function $N_{142}^{(g)}$ is defined in $T_7$. Fig. \ref{fig:nodal_edge_bubble} is generated by the script
\begin{verbatim}
  example4_draw_hp_basis_2D_global
\end{verbatim}
\end{example}

\subsection{Global indexing}
The relation between the topology of $\mathcal{T}$ and the global shape function indices is represented by three essential matrices. 
\subsubsection{A matrix $\hpindx$}
is of size $\nb \times 5$ and stores the key attributes of the global shape functions $\Nmg \in \SpT$, $1 \leq \m \leq \nb$ which are uniquely determined by:
the degree of $\Nmg$ (the first column of $\hpindx$),
the type of $\Nmg$ (nodal, edge or bubble) specified by the global index of the respective node (the $2$nd column), edge (the $3$rd column), or element (the $4$th column).
Additionally, the type of bubble requires a local index of a bubble (the $5$th column).
The key advantage of this approach is that for the same $\mathcal{T}$ and $1 \leq p_1 < p_2$ the first $n_{p_1}$ rows of both matrices $B(\mathcal{T},p_1)$ and $B(\mathcal{T},p_2)$ are the same.

\subsubsection{A matrix $\indx$} of size $\nbref \times \nt$ collects for individual elements the indices of the corresponding global functions. In particular, $C_{l,k}(\mathcal{T},\p) = \m$ means that $\Nmg\big|_{T_k}$ corresponds to the $l$-th local shape function on the $k$-th element.

\subsubsection{A matrix $\signs$} of size $\nbref \times \nt$ for the $l$-th row and the $k$-th column returns the sign of the $l$-th local function on the $k$-th element. For edges adjacent to two elements, the corresponding local edge functions of odd degrees have to be assigned opposite signs to ensure the continuity of the corresponding global edge functions.

\begin{table}[h]
    \centering
    \resizebox{0.85\columnwidth}{!}{
    \begin{tabularx}{\textwidth}
    {C |C |C |C |C }
   $\p$ & the global node index  & the global edge index & the global element index & the local bubble index \\
     \hline 
1 & 1 & - & - & - \\
 1 & 2 & - & - & - \\
 1 & 3 & - & - & - \\
 1 & 4 & - & - & - \\
 \hline
 2 & - & 1 & - & - \\
 2 & - & 2 & - & - \\
 2 & - & 3 & - & - \\
 2 & - & 4 & - & - \\
  \hline
 3 & - & 1 & - & - \\
 3 & - & 2 & - & - \\
 3 & - & 3 & - & - \\
 3 & - & 4 & - & - \\
  \hline
 4 & - & 1 & - & - \\
 4 & - & 2 & - & - \\
 4 & - & 3 & - & - \\
 4 & - & 4 & - & - \\
 4 & - & - & 1 & 1 \\
  \hline
5 & - & 1 & - & - \\
5 & - & 2 & - & - \\
5 & - & 3 & - & - \\
5 & - & 4 & - & - \\
5 & - & - & 1 & 1 \\
5 & - & - & 1 & 2 \
    \end{tabularx}}
    \vspace{0.3cm}
    \caption{The matrix $\hpindx$ for $\mathcal{T}$ with $\nn = 4$, $\ned = 4$, $\nt = 1$ and $p = 5$. Zero values are  replaced by symbol $'-'$.}
    \vspace{-10mm}
    \label{tab:hp_indexing}
\end{table}

\begin{example}
We assume a rectangulation $\mathcal{T}$ with $\nn = 4$, $\ned = 4$, and $\nt = 1$. Tab. \ref{tab:hp_indexing} depicts the corresponding matrix $\hpindx$ for $p = 5$ that can be generated by the script
\begin{verbatim}
  example5_B_matrix
\end{verbatim}
\end{example}

\begin{figure}[H]
\centering
\vspace{-2cm}
\begin{minipage}[c]{0.45\textwidth}
\includegraphics[width=\textwidth]{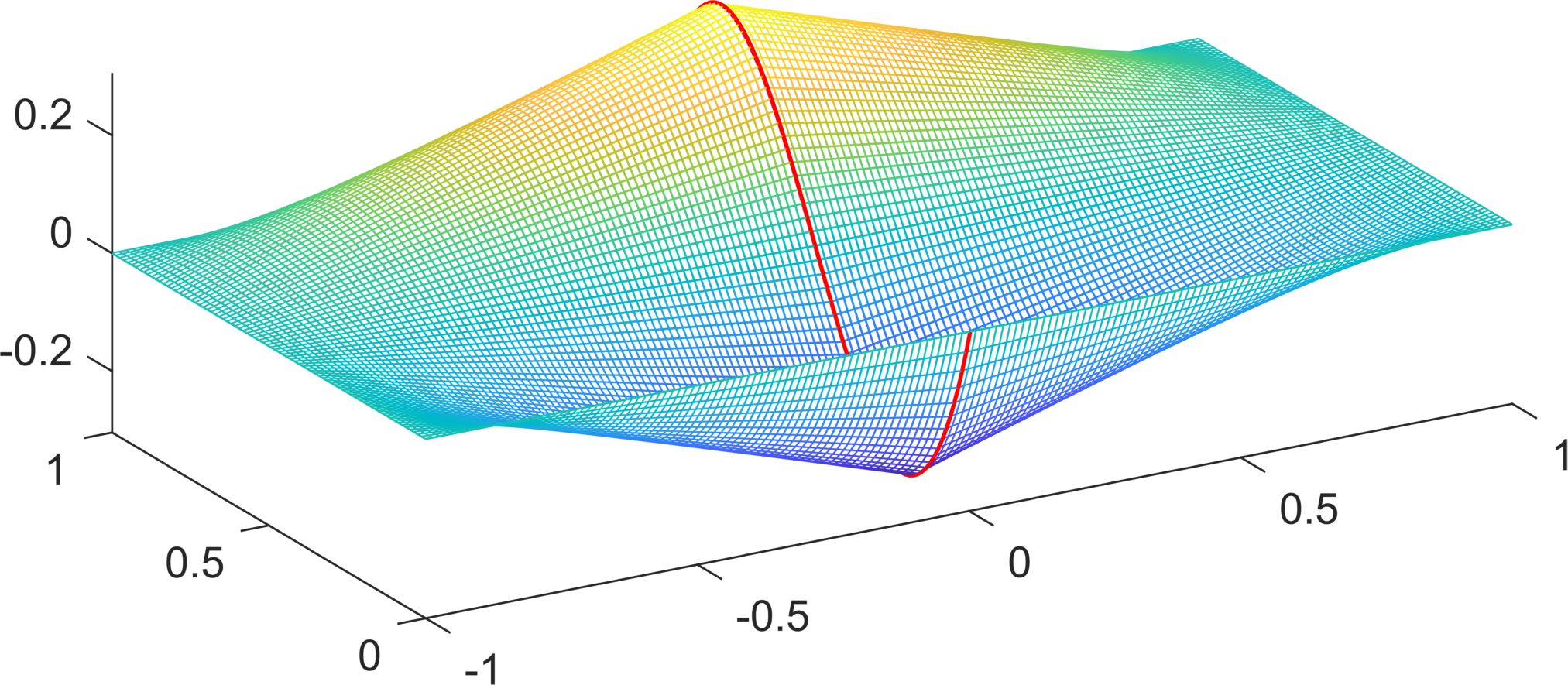}
\end{minipage}
\hspace{0.05\textwidth}
\begin{minipage}[c]{0.45\textwidth}
\includegraphics[width=\textwidth]{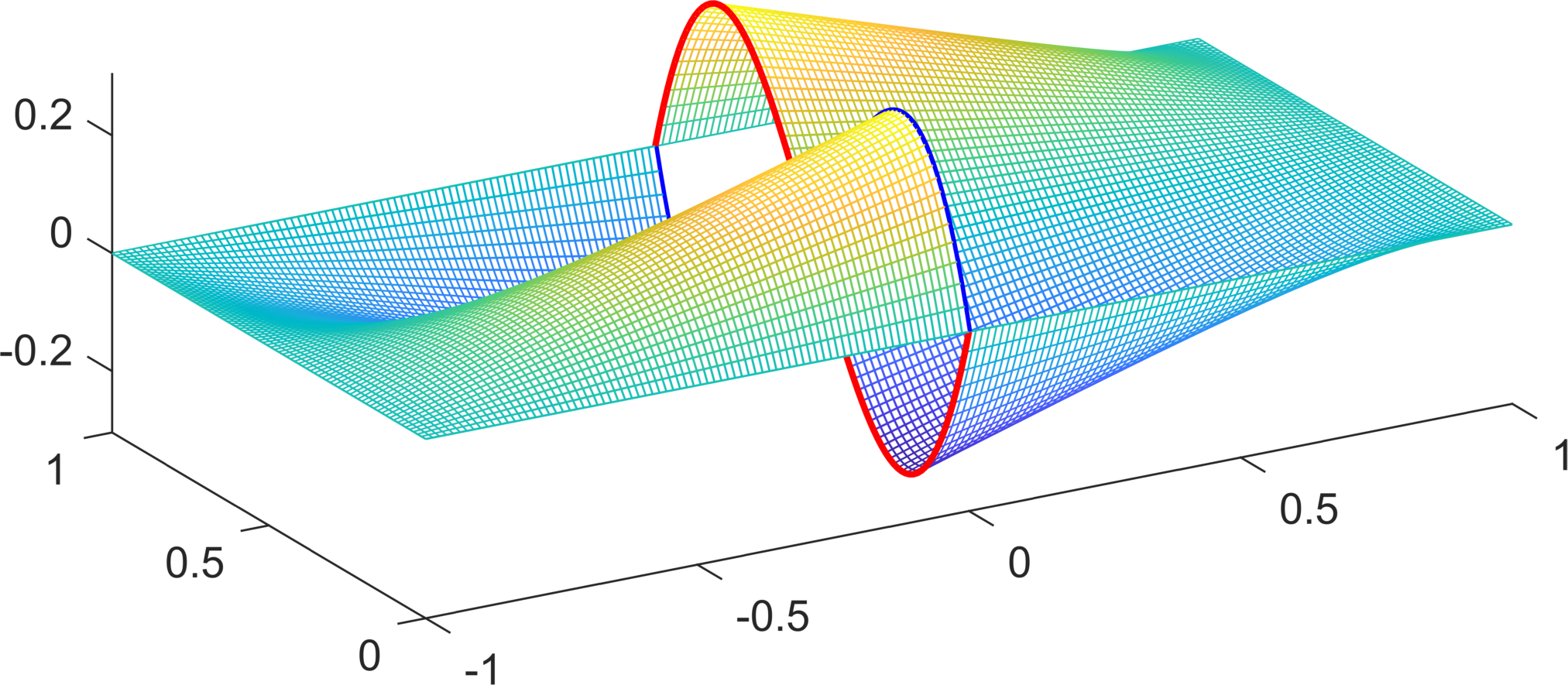}
\end{minipage} \:\:\:
\caption{The right (left) and wrong (right) orientation of $N_{17}^{(g)}$ of Example \ref{ex:6}.}
\label{fig:signs}
\end{figure}

\begin{example} \label{ex:6}
We assume a rectangulation $\mathcal{T}$ with $\nn = 6$, $\ned = 7$ and $\nt = 2$. Tab. \ref{tab:mesh_glob_indx} depict the corresponding matrices $\indx$ and $\signs$ for $p = 3$  which can be generated by the script
\begin{verbatim}
  example6_C_S_matrices
\end{verbatim}
Fig. \ref{fig:signs} shows the global edge function $N_{17}^{(g)}$ of the $3$rd degree. The left part exploits the right orientation with opposite signs providing continuity, and the right part exploits the wrong orientation leading to discontinuity.
\end{example}

\begin{table}[H]
    \centering
    \resizebox{0.9\columnwidth}{!}{
    \begin{tabularx}{0.45\textwidth}
    {C |C |C}
  $l$ & $T_1$ & $T_2$ \\
     \hline 
1 & 1 & 2 \\
2 & 2 & 3 \\
3 & 5 & 6 \\
4 & 4 & 5 \\
5 & 7 & 9 \\
6 & 10 & 11 \\
7 & 12 & 13 \\
8 & 8 & 10 \\
9 & 14 & 16 \\
10 & \textbf{17} & 18 \\
11 & 19 & 20 \\
12 & 15 & \textbf{17} \\
    \end{tabularx} \qquad \qquad
    \begin{tabularx}{0.45\textwidth}
    {C |C |C}
  $l$ & $T_1$ & $T_2$ \\
     \hline 
1 & 1 & 1 \\
2 & 1 & 1 \\
3 & 1 & 1 \\
4 & 1 & 1 \\
5 & 1 & 1 \\
6 & 1 & 1 \\
7 & 1 & 1 \\
8 & 1 & 1 \\
9 & 1 & 1 \\
10 & \textbf{-1} & 1 \\
11 & 1 & 1 \\
12 & 1 & \textbf{1} \\
    \end{tabularx}}
    \vspace{0.3cm}
    \caption{Matrices $\indx$ (left) and $\signs$ (right) of Example \ref{ex:6}.}
    \label{tab:mesh_glob_indx}
\end{table}

\section{Mass and stiffness matrices} \label{sec:4}

\subsection{The reference mass and stiffness matrices} 
are for a given $p \in \N$ matrices of size $\nbref \times \nbref$ defined by
\begin{equation}
    \Mref_{i,j} = \int_{\Tr} N_i N_j \dxdy \, , \qquad \Kref_{i,j} = \int_{\Tr} \nabla N_i \cdot \nabla N_j \dxdy \, .
\end{equation}
Functions
\begin{verbatim}
  mass_matrixQp_2D_reference(p)
    
  stiffness_matrixQp_2D_reference(p)
\end{verbatim}
evaluate the corresponding reference mass and stiffness matrices using the Gaussian quadrature rule. For a given $\p$, the function \verb+[X,W] = intrec_hp(p)+
returns the Gauss points $X \in \Tr$ together with the corresponding weights stored in a vector $W$.

\subsection{The global mass and stiffness matrices} \label{subsec:M}

are for a specific $p$ and $\mathcal{T}$ matrices of size $\nb \times \nb$ defined by
\begin{equation}
    M_{i,j} = \int_{\mathcal{T}} \Nig \Njg \dxdy \, , \qquad K_{i,j} = \int_{\mathcal{T}} \nabla \Nig \cdot \nabla \Njg \dxdy
\end{equation}
and assembled by adding the contributions of local mass and stiffness matrices $\Mk$ and $\Kk$ of size $\nbref \times \nbref$ to the corresponding entries. In particular, $M_{i,j}(T_k)$ and $K_{i,j}(T_k)$ contributes to the $\boldsymbol{c}_i^k$-th row and the $\boldsymbol{c}_j^k$-th column of $M$ and $K$, respectively, where $\boldsymbol{c}^k$ is the $k$-th column vector of $\indx$.

For any $T_k \in \mathcal{T}, \, 1 \leq k \leq \nt$ the local mass matrix $\Mk$ is given by
\begin{equation} \label{mass_loc}
    \Mk = \frac{|T_k|}{|T_{ref}|} \Mref = \frac{|T_k|}{4} \Mref \, ,
\end{equation}
however, this formula cannot be applied to the assembly of the local stiffness matrix $K(T_k)$. Instead, we apply the chain rule to evaluate
\begin{equation}
    \begin{split}
        K_{i,j}(T_k) &= \int_{T_k} \nabla \tilde{N}_i(x,y) \cdot \nabla \tilde{N}_j(x,y) \dxdy = \\
        &= \int_{T_k} \nabla N_i\big(Q^{-1}(x,y)\big) \cdot \nabla N_j\big(Q^{-1}(x,y)\big) \dxdy \, ,
    \end{split}
\end{equation}
where $\tilde{N}_i$ and $N_i$, $1 \leq i \leq \nbref$ are the $i$-th local function on $T_k$ and $\Tr$, respectively. Using the chain rule, one can write
\begin{equation}
    \begin{split}
        \nabla \tilde{N}_i\big(Q^{-1}(x,y)\big) = \Big(\frac{\partial \tilde{N}_i}{\partial \xi} \frac{\partial \xi}{\partial x} + \frac{\partial \tilde{N}_i}{\partial \eta} \frac{\partial \eta}{\partial x}, \; \frac{\partial \tilde{N}_i}{\partial \xi} \frac{\partial \xi}{\partial y} + \frac{\partial \tilde{N}_i}{\partial \eta} \frac{\partial \eta}{\partial y}\Big) \, ,
    \end{split}
\end{equation}
where
$$ \Big(\frac{\partial \xi}{\partial x}, \frac{\partial \eta}{\partial x}\Big) = \frac{\partial Q^{-1}}{\partial x}(x,y) \, , \quad \Big(\frac{\partial \xi}{\partial y}, \frac{\partial \eta}{\partial y}\Big) = \frac{\partial Q^{-1}}{\partial y}(x,y).$$
Additionally, we apply derivative of the formula of inverse function to evaluate
\begin{equation}
    \begin{pmatrix}
    \frac{\partial \xi}{\partial x} & \frac{\partial \xi}{\partial y} \\
    \frac{\partial \eta}{\partial x} & \frac{\partial \eta}{\partial y}
    \end{pmatrix}
    = \nabla Q^{-1}(x,y) = \big(\nabla Q(\xi,\eta)\big)^{-1} =
    \begin{pmatrix}
    \frac{\partial x}{\partial \xi} & \frac{\partial x}{\partial \eta} \\
    \frac{\partial y}{\partial \xi} & \frac{\partial y}{\partial \eta}
    \end{pmatrix}^{-1} \, .
\end{equation}
\vspace{-7mm}

\begin{table}[H]
    \centering
    \resizebox{0.95\columnwidth}{!}{
    \begin{tabularx}{0.6\textwidth}
    {C |C |C |C |C |C}
  level & $n_1$ & $n_2$ & $n_3$ & $n_4$ & $n_5$ \\
     \hline
2 & $2.5 \cdot 10^1$ & $6.5 \cdot 10^1$ & $1.1 \cdot 10^2$ & $1.6 \cdot 10^2$ & $2.3 \cdot 10^2$ \\
3 & $8.1 \cdot 10^1$ & $2.3 \cdot 10^2$ & $3.7 \cdot 10^2$ & $5.8 \cdot 10^2$ & $8.5 \cdot 10^2$ \\
4 & $2.9 \cdot 10^2$ & $8.3 \cdot 10^2$ & $1.4 \cdot 10^3$ & $2.2 \cdot 10^3$ & $3.2 \cdot 10^3$ \\
5 & $1.1 \cdot 10^3$ & $3.2 \cdot 10^3$ & $5.3 \cdot 10^3$ & $8.4 \cdot 10^3$ & $1.3 \cdot 10^4$ \\
6 & $4.2 \cdot 10^3$ & $1.3 \cdot 10^4$ & $2.1 \cdot 10^4$ & $3.3 \cdot 10^4$ & $5.0 \cdot 10^4$ \\
7 & $1.7 \cdot 10^4$ & $5.0 \cdot 10^4$ & $8.3 \cdot 10^4$ & $1.3 \cdot 10^5$ & $2.0 \cdot 10^5$ \\
8 & $6.6 \cdot 10^4$ & $2.0 \cdot 10^5$ & $3.3 \cdot 10^5$ & $5.3 \cdot 10^5$ & $7.9 \cdot 10^5$ \\
9 & $2.6 \cdot 10^5$ & $7.9 \cdot 10^5$ & $1.3 \cdot 10^6$ & $2.1 \cdot 10^6$ & $3.2 \cdot 10^6$ \\
    \end{tabularx} \qquad
    \begin{tabularx}{0.4\textwidth}
    {C |C |C |C}
  level & $\nn$ & $\ned$ & $\nt$ \\
     \hline
2 & $2.5 \cdot 10^1$ & $4.0 \cdot 10^1$ & $1.6 \cdot 10^1$ \\
3 & $8.1 \cdot 10^1$ & $1.4 \cdot 10^2$ & $6.4 \cdot 10^1$ \\
4 & $2.9 \cdot 10^2$ & $5.4 \cdot 10^2$ & $2.6 \cdot 10^2$ \\
5 & $1.1 \cdot 10^3$ & $2.1 \cdot 10^3$ & $1.0 \cdot 10^3$ \\
6 & $4.2 \cdot 10^3$ & $8.3 \cdot 10^3$ & $4.1 \cdot 10^3$ \\
7 & $1.7 \cdot 10^4$ & $3.3 \cdot 10^4$ & $1.6 \cdot 10^4$ \\
8 & $6.6 \cdot 10^4$ & $1.3 \cdot 10^5$ & $6.6 \cdot 10^4$ \\
9 & $2.6 \cdot 10^5$ & $5.3 \cdot 10^5$ & $2.6 \cdot 10^5$ \\
    \end{tabularx}}
    \vspace{0.2cm}
    \caption{The numbers of global shape functions (left) and mesh properties (right) of Example \ref{example7}.} \label{tab:mesh_properties_MK}
\centering
\resizebox{0.9\columnwidth}{!}{
\begin{tabularx}{\textwidth}{C | C C | C C| C C| C C| C C}
& \multicolumn{2}{c|}{$\p = 1$}
& \multicolumn{2}{c|}{$\p = 2$}
& \multicolumn{2}{c|}{$\p = 3$}
& \multicolumn{2}{c|}{$\p = 4$}
& \multicolumn{2}{c}{$\p = 5$}
\\
\hline
level
& M [s] & K [s]
& M [s] & K [s]
& M [s] & K [s]
& M [s] & K [s]
& M [s] & K [s]
\\
\hline
2 & 0.01 & 0.02 & 0.01 & 0.01 & 0.00 & 0.00 & 0.00 & 0.01 & 0.01 & 0.01 \\
3 & 0.00 & 0.01 & 0.00 & 0.00 & 0.00 & 0.01 & 0.01 & 0.01 & 0.01 & 0.01 \\
4 & 0.00 & 0.01 & 0.00 & 0.00 & 0.01 & 0.02 & 0.01 & 0.02 & 0.01 & 0.03 \\
5 & 0.00 & 0.01 & 0.00 & 0.01 & 0.01 & 0.02 & 0.01 & 0.05 & 0.02 & 0.09 \\
6 & 0.00 & 0.02 & 0.01 & 0.04 & 0.02 & 0.08 & 0.03 & 0.16 & 0.07 & 0.33 \\
7 & 0.01 & 0.08 & 0.03 & 0.15 & 0.07 & 0.32 & 0.25 & 0.71 & 0.45 & 1.46 \\
8 & 0.04 & 0.31 & 0.15 & 0.66 & 0.48 & 1.44 & 0.98 & 3.14 & 2.00 & 6.20 \\
9 & 0.22 & 1.29 & 0.91 & 2.86 & 1.94 & 6.01 & 5.26 & 12.43 & 11.17 & 27.31 \\
    \end{tabularx}}
    \vspace{0.2cm}
    \caption{Assembly times of mass and stiffness matrices in Example \ref{example7}.} \label{tab:times_MK}
\end{table}
\vspace{-12mm}
\begin{example}\label{example7}
For $\Omega = \Tr = [-1,1]^2$ the script
\begin{verbatim}
  example7_M_K_matrices_times
\end{verbatim}
runs a nested loop on different $\p$ and levels of uniform refinements of $\Omega$. The mass and stiffness matrices are assembled by the functions \verb+mass_matrixQp_2D(mesh)+ and \verb+stiffness_matrixQp_2D(mesh)+, respectively. Tables \ref{tab:mesh_properties_MK} and \ref{tab:times_MK} contain the properties of the mesh and the corresponding assembly times.
\end{example}
\vspace{-3mm}

\section{Solving partial differential equation in 2D}  \label{sec:5}
We solve a diffusion-reaction boundary value problem
\begin{equation} \label{drbvp}
\begin{split}
    -\Delta u + \nu \, u = f \quad \mbox{in }  \Omega \, , \qquad \frac{\partial u}{\partial n} = 0 \quad \mbox{on } \partial \Omega
\end{split}
\end{equation}
by applying the hp-FEM method to the weak formulation of \eqref{drbvp} given by
\begin{equation} \label{drbvp_fem}
   \int_{\mathcal{T}} \nabla u \cdot \nabla \Nmg \dxdy \, + \, \nu \int_{\mathcal{T}} u \, \Nmg \dxdy = \int_{\mathcal{T}} f \, \Nmg \dxdy \, , \quad \forall \, \Nmg \in \SpT \, .
\end{equation}
It leads to the algebraic system of linear equations in the form of
\begin{equation} \label{linsystem}
    (K + \nu \, M)\, \tilde{u}_n = b \, ,
\end{equation}
where $M$ and $K$ are global mass and stiffness matrices, $u_n$ is the numerical solution of \eqref{drbvp_fem} represented by vector $\tilde{u}_n \in \R^{\nb}$ of coefficients in the corresponding hp basis and vector $b \in \R^{\nb}$ is given by $ b_m = \int_{\mathcal{T}} f N_m^{(g)} \dxdy$.
We assume the domain $\Omega = \Tr = [-1,1]^2$ and the parameter $\nu = 0.1$. It is easy to show that 
\begin{equation*}
    u(x,y) = (1-x^2)^2 \, (1-y^2)^2
\end{equation*}
represents the solution of \eqref{drbvp}
corresponding to the function
$$f(x,y) = \nu \, u(x,y) -4\big( -2 +5y^2 -y^4 +x^4(-1 +3y^2) +x^2(5 -12y^2 +3y^4) \big)$$ for $(x,y) \in \Omega$.
To study the convergence of hp-approximations, we take several levels of uniform refinements of $\Omega$ and solve \eqref{linsystem} for different $\p$, $1 \leq p \leq p_{max}$. The exact solution $u$ is approximated in $S^{\tilde{p}}(\mathcal{T})$, $\tilde{p} = p_{max}+2$ by vector $\tilde{u}$. The corresponding error $e_n$ in the energy norm is given by
\begin{equation} \label{energyerror}
    (e_n)^2  = \int_{\mathcal{T}} \big(\|\nabla u - \nabla u_n\|^2 + (u-u_n)  ^2\big) \dxdy \, \approx \,  (\tilde{u}-\tilde{u}_n)^T (K+M) (\tilde{u}-\tilde{u}_n) \, .
\end{equation}
\vspace{-7mm}

\begin{figure}[H]
\begin{minipage}[c]{0.3\textwidth}
\includegraphics[width=\textwidth]{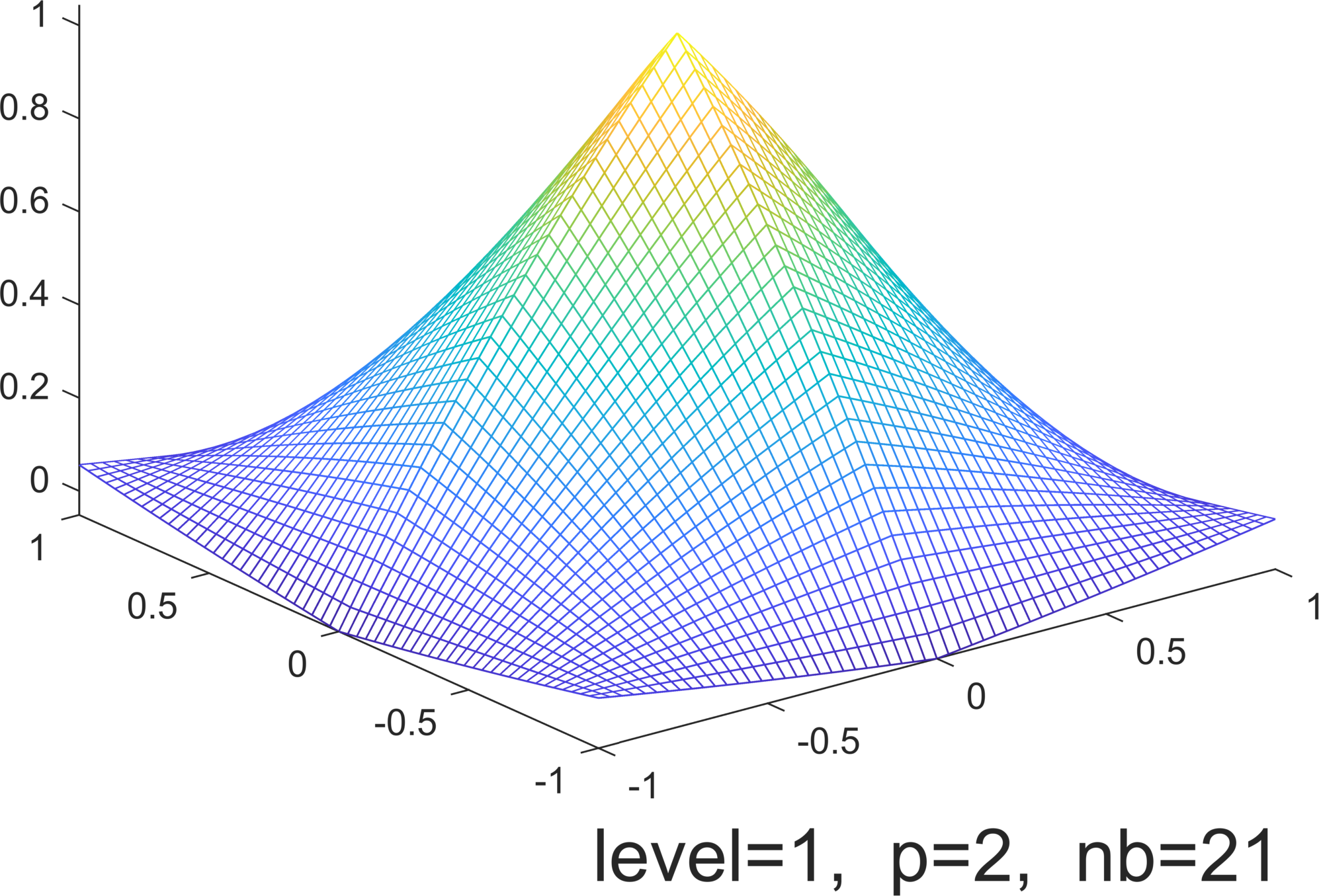} \\
\vspace{0.1cm}
\includegraphics[width=\textwidth]{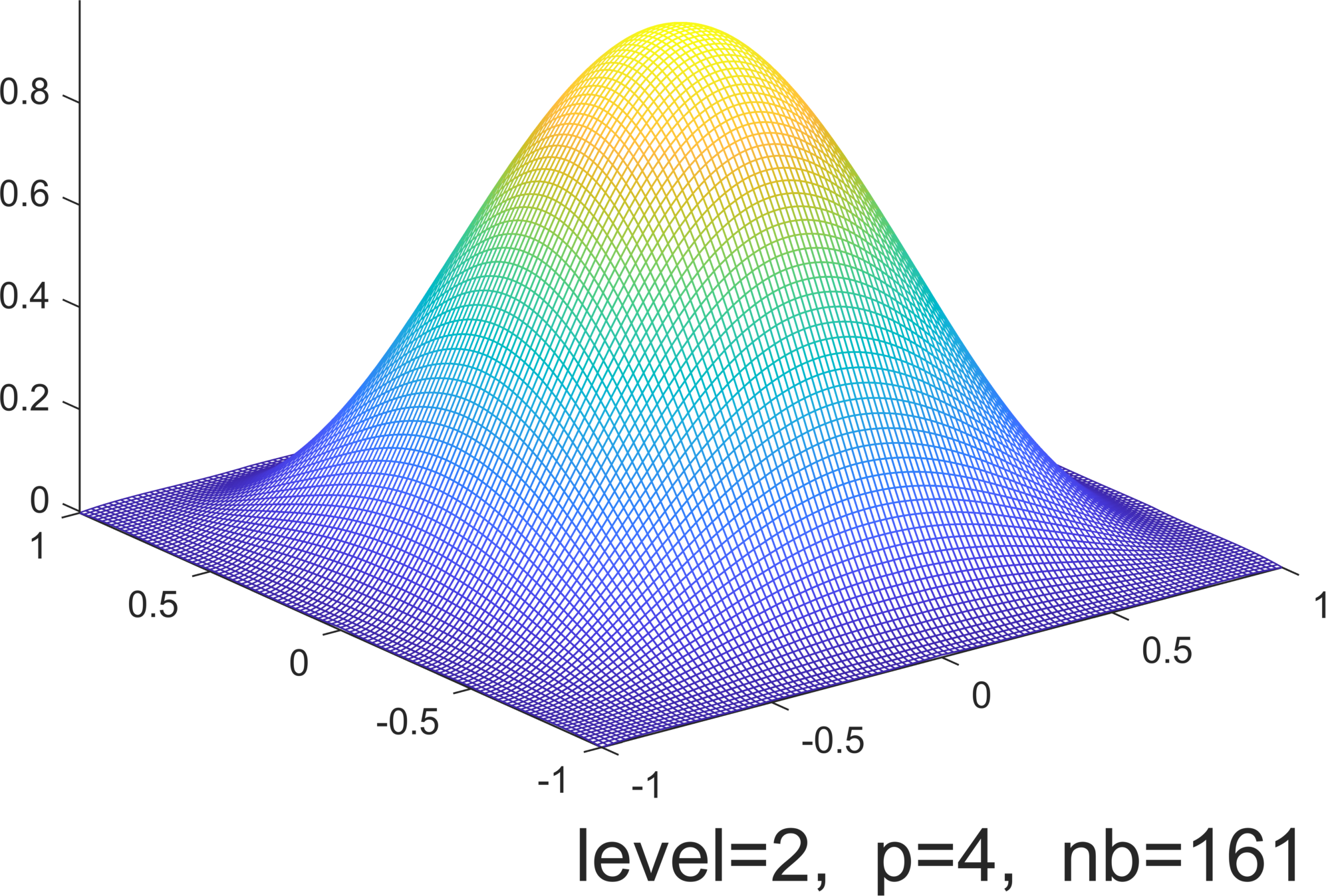}
\end{minipage} \;\;
\begin{minipage}[c]{0.65\textwidth}
\includegraphics[width=\textwidth]{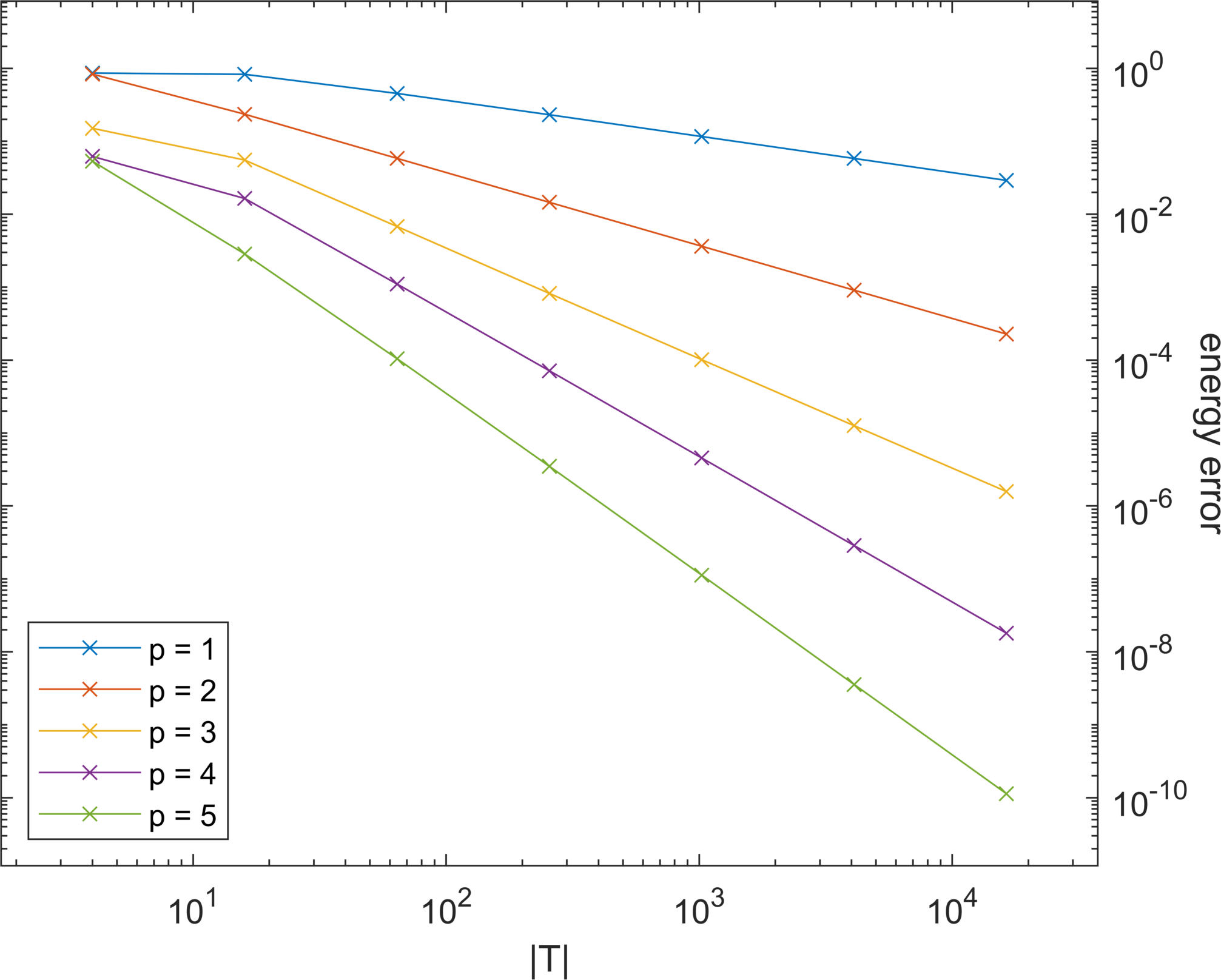}
\end{minipage}
\caption{Examples of solutions of \eqref{drbvp_fem} and convergence in energy norm.}
\label{fig:solutions}
\end{figure}
\vspace{-5mm}

\noindent
The script
\begin{verbatim}
  example8_diffusion_reaction_BVP
\end{verbatim}
utilizes a nested for loop on $\p$ (inside) and mesh refinement levels (outside). Two particular numerical solutions are shown in Fig. \ref{fig:solutions} (left). The corresponding errors \eqref{energyerror} are depicted in Fig. \ref{fig:solutions} (right).

\subsection{Implementation remarks 
}
Assembly times in Tab. \ref{tab:times_MK} were obtained on a MacBook Air (M1 processor, 2020) with 16 GB memory running MATLAB R2022a. Complementary software for this paper is available at 
\begin{center}
\url{https://www.mathworks.com/matlabcentral/fileexchange/111420} 
\end{center}
for download and testing. The codes for the evaluation of the shape functions were provided by Dr. Sanjib Kumar Acharya (Mumbai). The assemblies of FEM matrices are based on our vectorized codes \cite{AnjamValdman2015, RahmanValdman2013}. The names of most of the mesh attributes and the domain rectangulation algorithms are taken from \cite{AMJV}. 



\bibliographystyle{splncs03}

\end{document}